\documentclass[%
 aip,
 amsmath,amssymb,
 reprint,%
]{revtex4-1}

\usepackage{graphicx}
\usepackage{dcolumn}
\usepackage{bm}

\usepackage[utf8]{inputenc}
\usepackage[T1]{fontenc}
\usepackage{mathptmx}
\usepackage{textcomp}
\usepackage{graphicx}
\usepackage{setspace}
\usepackage{epsfig}
\usepackage{natbib}
\usepackage{pifont}
\usepackage{braket}
\usepackage{amssymb} 
\usepackage{mathtools}
\usepackage{graphicx}
\usepackage{textcomp}
\usepackage{amsmath}
\usepackage{geometry}
\usepackage[normalem]{ulem}
\usepackage{color}
\usepackage{float}
\usepackage{gensymb}
\usepackage{multirow}
\usepackage{soul}

\newcommand{\beginsupplement}{%
        \setcounter{table}{0}
        \renewcommand{\thetable}{S\arabic{table}}%
        \setcounter{figure}{0}
        \renewcommand{\thefigure}{S\arabic{figure}}%
     }

\begin{document}


\title{Reconstructive Spectrometer using Photonic Crystal Cavity}

\author{Naresh Sharma}
\affiliation{Department of Electrical Engineering, Indian Institute of Technology Kanpur, Kanpur-208016, UP, India}
\author{Govind Kumar}%
\affiliation{The Centre for Lasers and Photonics, Indian Institute of Technology Kanpur, Kanpur-208016, UP, India} 
\author{Vivek Garg}
\affiliation{Department of Mechanical Engineering, Indian Institute of Technology  Bombay, Mumbai-400076, Maharashtra, India}
\author{Rakesh G. Mote}
\affiliation{Department of Mechanical Engineering, Indian Institute of Technology  Bombay, Mumbai-400076, Maharashtra, India}
\author{Shilpi Gupta}
\email{ShilpiG@iitk.ac.in}
\affiliation{Department of Electrical Engineering, Indian Institute of Technology Kanpur, Kanpur-208016, UP, India}

\date{\today}

\begin{abstract}

Optical spectrometers have propelled scientific and technological advancements in a wide range of fields. 
While sophisticated systems with excellent performance metrics are serving well in controlled laboratory environments, many applications require systems that are portable, economical, and  robust to optical misalignment. Here, we propose and demonstrate a spectrometer that uses a planar one-dimensional photonic crystal cavity as a dispersive element and a reconstructive computational algorithm to extract spectral information from spatial patterns. The simple fabrication and planar architecture of the photonic crystal cavity render our spectrometry platform economical and robust to optical misalignment. The reconstructive algorithm allows miniaturization and portability. The intensity transmitted by the photonic crystal cavity has a wavelength-dependent spatial profile. We generate the spatial transmittance function of the system using finite-difference time-domain method and also estimate the dispersion relation. The transmittance function serves as a transfer function in our reconstructive algorithm. We show accurate estimation of various kinds of input spectra. We also show that the spectral resolution of the system depends on the cavity linewidth that can be improved by increasing the number of periodic layers in distributed Bragg mirrors. Finally, we experimentally estimate the center wavelength and linewidth of the spectrum of an unknown light emitting diode. The estimated values are in good agreement with the values measured using a commercial spectrometer.

\end{abstract}

\maketitle

Optical spectrometers are essential tools for scientific and industrial applications across a wide range of fields, including materials science \cite{Richard2018}, biological and chemical sensing \cite{bacon2004, cai2017}, environmental science  \cite{bacon2004}, and nanotechnology \cite{adams2013}. The traditional approach for designing spectrometers involves dispersion of light using an optical element such as a grating, propagation along large optical path lengths to reach a detector, and point-by-point mapping from spatial to spectral domain. Though these spectrometers offer high resolution, they tend to be bulky and expensive that hinders their usage in applications requiring portability and wide deployment. Alternate design approach has been to use tuneable narrowband filters in which different bands of the spectrum are measured by either tuning the spectral response of the filter over time \cite{lammel2001, yao2019, nitkowski2008} or by using an array of detectors for an array of filters \cite{meng2014, xia2011, kyotoku2010, pervez2010,wang2007}. While these spectrometers are compact, they require complex and expensive fabrication techniques. Both these approaches are also sensitive to optical misalignment.

\begin{figure}[t]
\centering\includegraphics[width=6.5 cm]{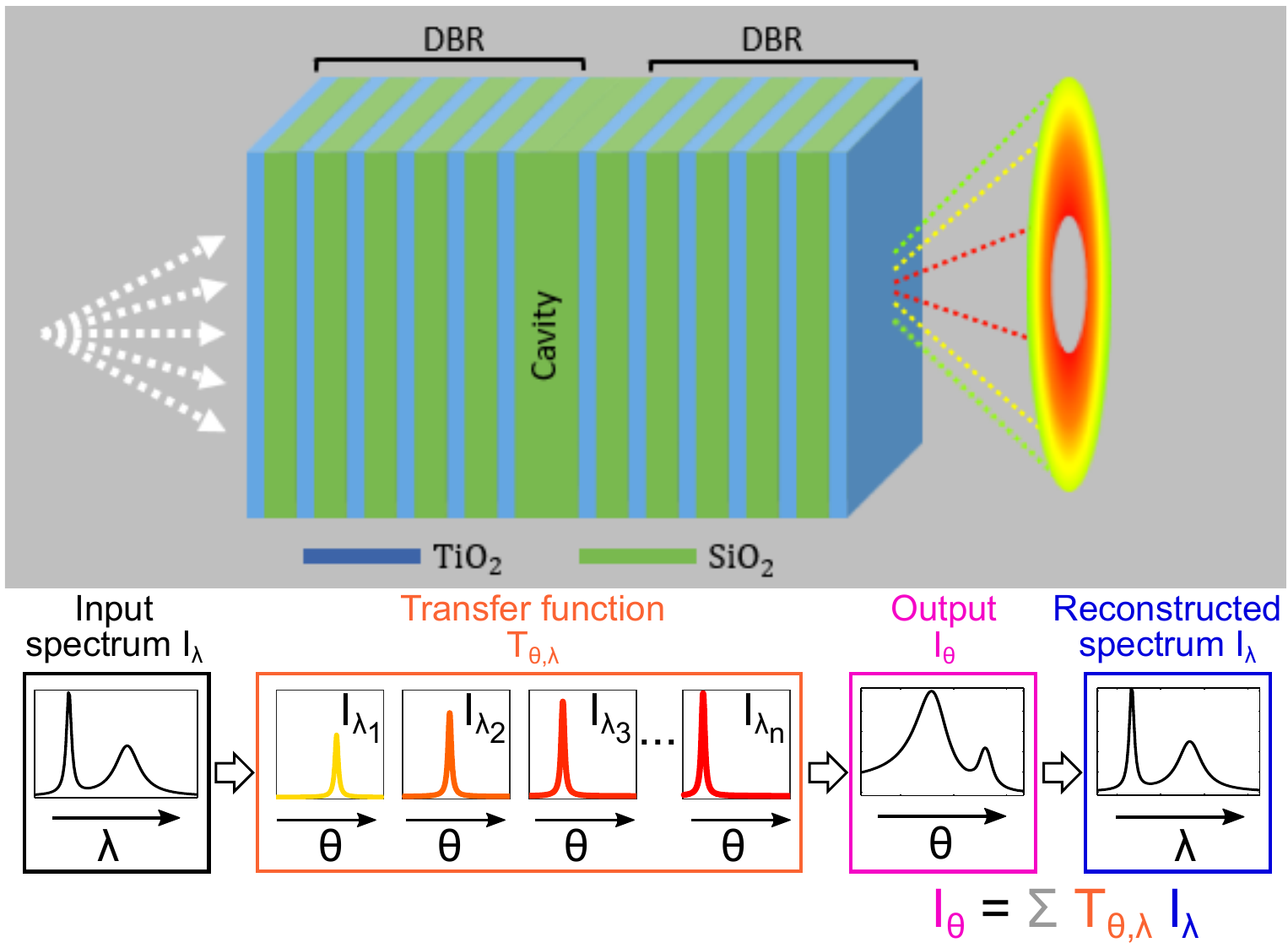}
\caption{\textbf{Working principle}: A reconstructive spectrometer employing a one-dimensional photonic crystal cavity. The cavity exhibits wavelength-dependent spatial profile of the transmitted intensity, which is used in the spectral reconstruction algorithm.}
\label{Fig:Fig1_Schematic}
\end{figure}

To eliminate the need for point-by-point mapping from spatial to spectral domain, and thus, long path-lengths to achieve high resolution, computational reconstruction techniques have been employed for spectrum retrieval from pre-calibrated detector responses \cite{Yang2021}. These techniques take advantage of the unique signature patterns generated by the incident wavelengths in the spatial domain after passing through a dispersive element or a filter and reconstruct the incident light spectrum through spatial-spectral mapping by solving a set of linear equations \cite{Yang2021}. Many different platforms such as disordered photonic chip \cite{redding2013}, evanescently coupled spiral waveguide \cite{Redding2016}, dispersive hole array \cite{Yang2015}, polychromator \cite{wang2014}, and colloidal quantum dots \cite{bao2015} have been used to demonstrate these reconstructive spectrometers. While these platforms have shown promising performance, most of them still require complex and expensive fabrication techniques and are sensitive to optical misalignment, hindering mass-production for wide deployment. Therefore, there is a need to design spectrometers that are portable, economical, and  robust to optical misalignment.

Here, we propose and demonstrate a reconstructive spectrometer that uses a planar one-dimensional photonic crystal cavity as a dispersive element to achieve miniaturization, cost-effectiveness, and robustness to misalignment. The photonic crystal cavity acts as a spatial and spectral filter simultaneously. It spatially separates different spectral components present in the input beam into annular beams of different divergence angles \cite{sharma2020} --- a signature pattern that we use for mapping from spatial to spectral domain (Figure \ref{Fig:Fig1_Schematic}). Further, the planar architecture and in-plane symmetry of the photonic crystal cavity make our spectroscopic platform robust to optical misalignment due to translational invariance \cite{sharma2020}. Therefore, our platform is portable, economical, easy-to-use, and easy to mass-produce. In this work, we determine spatial transmittance function of the system using finite-difference time-domain (FDTD) simulations, which provides a map between the divergence angle and wavelength of the transmitted beam. This map acts as a transfer function of our system, which we use to reconstruct the input spectrum. We show that this technique can be used to accurately estimate various kinds of input spectrum. Using least square error method, 
we show the robustness of our technique by estimating spectrum in the presence of noise. The spectral resolution of our system is governed by the cavity linewidth and the spectral range is governed by the stopband. Finally, as a proof-of-concept demonstration, we estimate the center wavelength and linewidth of a light emitting diode (LED) using a fabricated and characterized photonic crystal cavity and find a good agreement with measurements obtained using a commercial spectrometer.

\section*{System Design}

\begin{figure}[b]
\centering\includegraphics[width=7 cm]{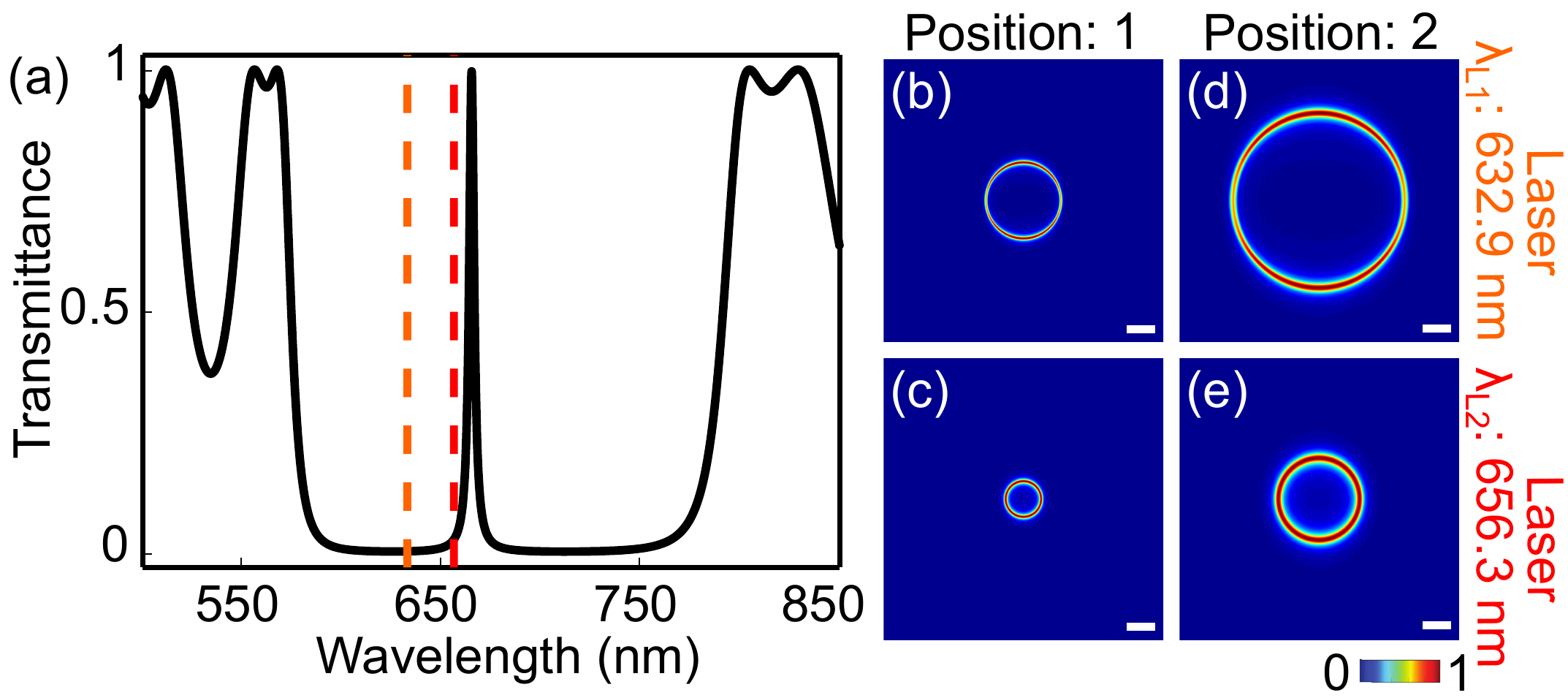}
\caption{\textbf{Transmitted intensity profiles for linearly polarized incident laser beams}: (a) Simulated transmittance spectrum for the designed photonic crystal cavity at normal incidence. The vertical dashed lines represent the laser wavelengths, 656.3 nm and 632.9 nm, used for characterization. Simulated far-field intensity profiles of laser beam after passing through the sample at (b)-(c) position 1 and (d)-(e) position 2, which are separated by 33 mm along the normal of the sample. All scale bars are 10 mm. Intensity of each panel is individually normalized.}
\label{Fig:Fig2_System_Design}
\end{figure}

A one-dimensional photonic crystal cavity is composed of two distributed Bragg reflectors (DBR) and a cavity spacer (Figure \ref{Fig:Fig1_Schematic}). We design the DBR as a stack of six periodic layers of silicon dioxide (SiO$_2$) and titanium dioxide (TiO$_2$) with quarter-wavelength optical thicknesses. The cavity spacer is a SiO$_2$ thin film of half-wavelength optical thickness. We calculate transmission spectrum of the designed cavity using FDTD method (Lumerical Inc) by illuminating with a plane-wave source at normal incidence (Figure  \ref{Fig:Fig2_System_Design}a). The cavity resonance wavelength is 665 nm and the stopband ranges from 575 nm to 790 nm. 

We simulate the far-field transmitted-intensity profiles of the cavity for a linearly polarized laser beam (at wavelengths $\lambda_{L1}$ = 632.9 nm and $\lambda_{L2}$ = 656.3 nm) incident on the structure using a 0.9 numerical aperture (NA) thin lens (Figures \ref{Fig:Fig2_System_Design}b-c). A one-dimensional photonic crystal cavity transforms a Gaussian profile of the incident laser beam into an annular profile in transmission if the laser wavelength is shorter than the cavity resonance wavelength at normal incidence \cite{sharma2020}. This transformation happens because the incident wavelength satisfies the resonance condition, imposed by the cavity, at an oblique angle of incidence. A wide range of incidence angles is simultaneously generated by the high-NA lens before the structure, and the cavity transmits a narrow band of incidence angle (the center angle of this narrow band is denoted as the resonance angle for the incident wavelength). Every incident wavelength, which lies within the stopband and is shorter than the cavity resonance wavelength, transmits as an annular beam with a unique resonance angle --- a signature pattern in the spatial domain for every wavelength. 

The focused beam (after the objective lens) has TE nature in the horizontal direction and TM nature in the vertical direction \cite{mansuripur1986}, which results in different diameter of the transmitted annular beam in horizontal and vertical directions for a particular incident wavelength and at a fixed distance from the sample (Figures \ref{Fig:Fig2_System_Design}b-c)\cite{sharma2020}. The diameter of the transmitted annular beam depends on the resonance angle of the cavity at the incident wavelength, which is a function of the difference between the wavelengths of the incident spectrum and the cavity resonance ($\lambda_{c}$) at normal incidence \cite{sharma2020}. The width of the transmitted annular beam depends on the cavity linewidth. We calculate the resonance angle ($\theta$) and resonance width ($\delta\theta$, defined as full-width half maxima at the resonance angle) using the diameters and widths of the annular beam calculated at two different positions, position-1 (Figure \ref{Fig:Fig2_System_Design}b-c) and position-2 (Figures \ref{Fig:Fig2_System_Design}d-e) for TE and TM polarization directions. The calculated resonance angles (resonance widths) for our designed cavity for the laser wavelengths $\lambda_{L1} =$ 632.9 nm and $\lambda_{L2} =$ 656.3 nm are 30.06$^\circ$ (2.33$^\circ$ ) and 15.35$^\circ$ (3.05$^\circ$) for TM polarization and 29.85$^\circ$ (1.73$^\circ$) and 15.28$^\circ$ (2.62$^\circ$) for TE polarization, respectively.

\begin{figure}[t]
\centering\includegraphics[width=7cm]{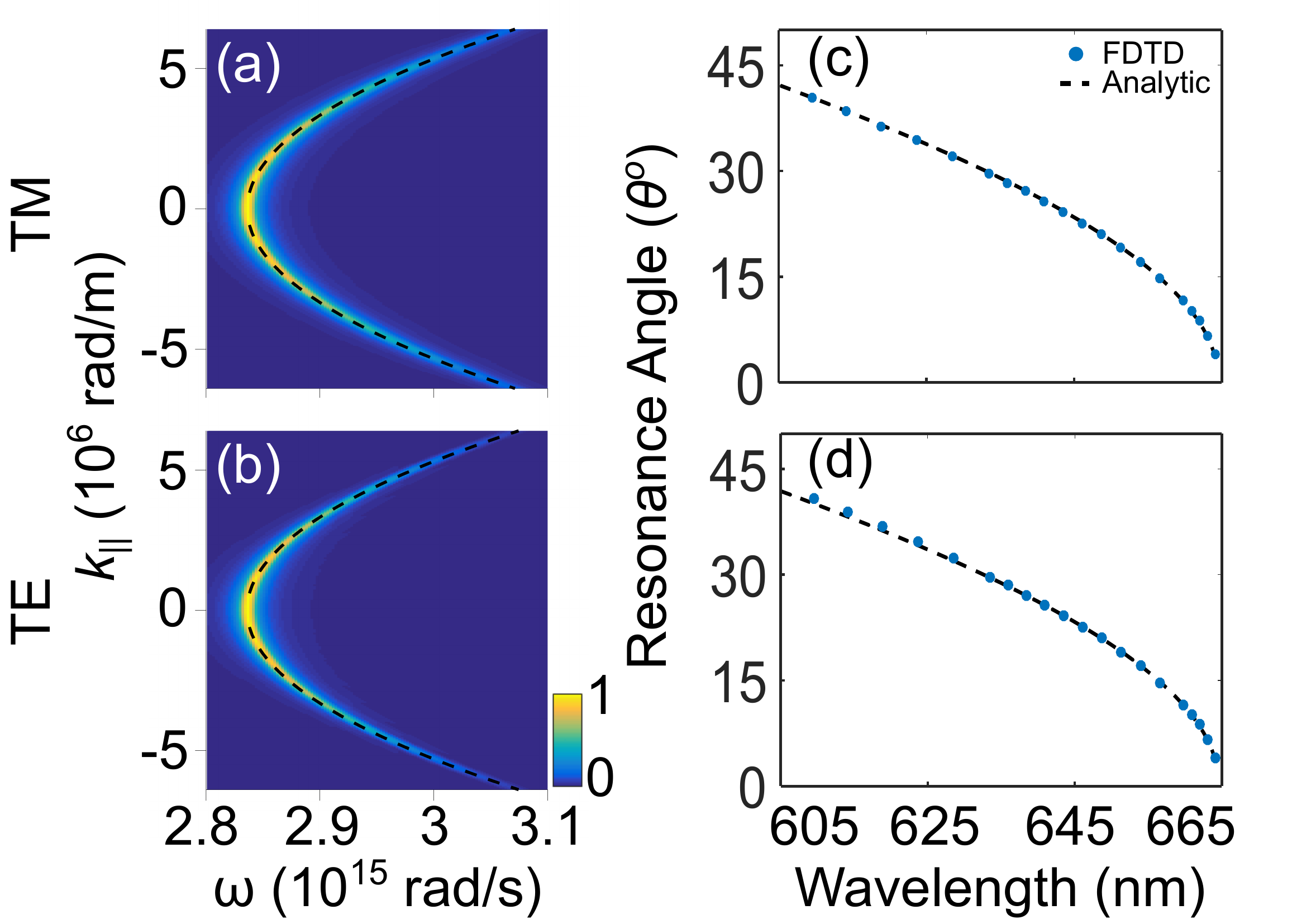}
\caption{\textbf{Transfer function and Dispersion Relation}: (a)-(b) Transfer function as a function of parallel component of the wave vector ($k_{||}$) and input frequency ($\omega$) for TM and TE polarization directions. The dashed curves represent analytic dispersion relation. (c)-(d) The resonance angle ($\theta$) as a function of input wavelength ($\lambda$) for TM and TE polarization directions. $k_{||}$ = $k_0$sin$\theta$. The markers represent FDTD simulation results and the dashed curves represent analytic results.} 
\label{Fig:Fig3_Fitting_Dispersion_relation}
\end{figure}
\textit{Transfer function and Dispersion Relation:} Following the same procedure as above, we simulate transmitted spatial intensity profile of a linearly polarized laser source of unit amplitude as a function of the source wavelength, which acts as the transfer function of the system (objective lens and photonic crystal cavity). We plot the transfer function as a function of the source frequency ($\omega$) and the in-plane component of the wavevector ($k_{||}$), which depends on the source wavelength and the resonance angle, for both TM and TE polarization directions (Figures \ref{Fig:Fig3_Fitting_Dispersion_relation}a-b). We observe that $\omega$ and $k_{||}$ of transmittance maxima exhibit a parabolic relationship. We model the dispersion relation between $\omega$ and $k_{||}$ using the following expression:

\begin{equation}
\label{eqn:Lambda_Theta}
\omega = A k_{||}^2 + \omega_c
\end{equation}
where $\omega$=2$\pi$c/$\lambda$, $\omega_c$=2$\pi$c/$\lambda_c$, $k_0$=2$\pi$/$\lambda$, $k_{||} = k_0sin\theta$, $\theta$ is the resonance angle at the incident wavelength $\lambda$, and $A$ is a constant which depends on the design parameters of the photonic crystal cavity and NA of the lens. We estimate $A$ and $\lambda_c$ as 5.74 (5.8) and 664.58 nm (664.63 nm) for TM (TE) polarization using the resonance angle  values (calculated in the previous section using the diameters of the annular beams) corresponding to $\lambda_{L1}$ and $\lambda_{L2}$. Using these estimated parameters, we plot the expression of the dispersion relation (eq. \ref{eqn:Lambda_Theta}) as black dashed curves in figures \ref{Fig:Fig3_Fitting_Dispersion_relation}a-b. The calculated dispersion relation using eq. \ref{eqn:Lambda_Theta} matches well with the transmittance maxima of the simulated transfer function obtained using the FDTD method. We also estimate the cavity linewidth at normal incidence as 2.86 nm (2.68 nm) for TM (TE) polarization using the resonance widths calculated in the previous section and eq. \ref{eqn:Lambda_Theta}. The estimated resonance wavelength 664.6 nm and the cavity linewidth  2.77 nm using the dispersion relation, averaged over both the polarization directions, show good agreement with the designed values (665 nm and 2.98 nm, Figure \ref{Fig:Fig2_System_Design}a). Our simple method shows that the dispersion relation of the system can be completely characterized (finding $A$, cavity resonance wavelength, and cavity linewidth) by measuring diameter of the transmitted annular beam at two positions for two known laser wavelengths. We further show that the bandedges of the stopband can also be characterized using the transmitted intensity profiles (Supplementary Figure \ref{SuppFig: design2and3}).

Figures \ref{Fig:Fig3_Fitting_Dispersion_relation}c-d show the variation of the resonance angle ($\theta$) with incident wavelength ($\lambda$) for TM and TE polarization directions. The dashed curves are calculated using eq. \ref{eqn:Lambda_Theta} and the markers are values estimated using the FDTD simulations. These plots are just another version of the dispersion curves, shown in figures \ref{Fig:Fig3_Fitting_Dispersion_relation}a-b, now in $\lambda-\theta$ space. Figures \ref{Fig:Fig3_Fitting_Dispersion_relation}c-d act as maps between the resonance angle, which can be easily estimated by measuring the diameter of the transmitted annular beam at two positions, and the wavelength of a narrowband spectrum being transmitted for a designed photonic crystal cavity.

\begin{figure}[t]
\centering\includegraphics[width=7 cm]{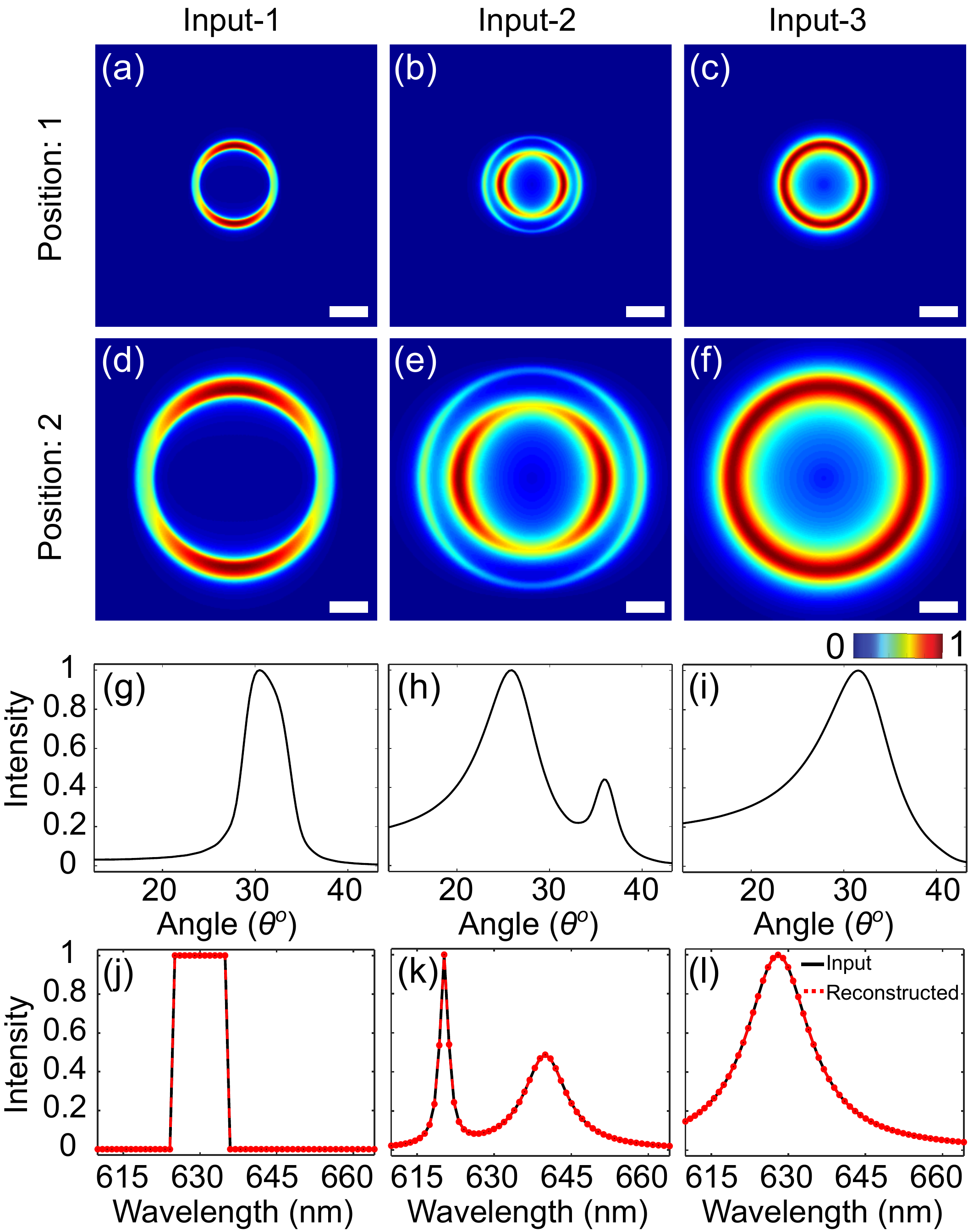}
\caption{\textbf{Reconstructing spectral intensity profile from spatial intensity profile}: Simulated far-field transmitted intensity profiles of three input spectra at (a-c) position 1 and (d-f) position 2. (g-f) Transmitted intensity profiles, averaged over TM and TE polarization directions, as a function of divergence angle. (j-l) Incident and reconstructed spectra of different inputs. The red dashed curves with circular markers represent the reconstructed spectra and the black solid curves represent the input spectra set in the simulation. The distance between on-axis position 1 and position 2 is 33 mm. All scale bars are 10 mm. Intensity of each panel is individually normalized.} 
\label{Fig:Figure4_LED_Intensity2Pos_simulation}
\end{figure}

\begin{table*}[t!]
\caption{Parameters of input spectra}
\centering
\begin{tabular}{c c c c c c}
\hline
 & Polarization & Profile & Peak wavelength  & Bandwidth  & Amplitude \\
\hline
Input-1 &  Vertical & Top Hat & 630 nm & 10 nm & 1 \\
\multirow{2}{*}{Input-2}& \multirow{2}{*}{ Horizontal} & \multirow{2}{*}{Two Lorentzian peaks} & 620 nm &  2 nm & 1 \\
&&& 640 nm & 10 nm & 0.5 \\
Input-3 & Unpolarized & One Lorentzian peak & 628 nm & 15 nm & 1 \\ 
\hline
\end{tabular}
\label{table:sources}
\end{table*}

Finally, we test our platform to estimate spectra of three different broadband input spectra (Table \ref{table:sources}). We simulate far-field transmitted intensity profiles for each input spectrum at two positions from the photonic crystal cavity (Figures \ref{Fig:Figure4_LED_Intensity2Pos_simulation}a-f). Using these far-field transmitted intensity profiles, we calculate transmitted intensity profiles as a function of the divergence angle, averaged over the TM and TE polarization directions (Figures  \ref{Fig:Figure4_LED_Intensity2Pos_simulation}g-i). The transmitted spatial intensity profile (I($\theta$)) is related to the incident spectral intensity profile (I($\lambda_n$)) by the following expression \cite{bao2015}:
\begin{equation}
\label{eqn:TransferFunction}
I(\theta) = \sum_{n=1}^{N} T(\theta,\lambda_n) * I(\lambda_n)
\end{equation}
where $T(\theta,\lambda_n)$ is the transfer function of the system (the photonic crystal cavity and the objective lens) and represents the transmitted spatial intensity profile of a laser source of unit amplitude and the center wavelength of $\lambda_n$. Figure \ref{SuppFig: Transferfunction} shows the transfer function matrix of our system calculated by using unpolarized laser sources of unit amplitude and different wavelengths. The solution of equation \ref{eqn:TransferFunction} is susceptible to measurement errors and instrument noise, and therefore, we calculate the optimal solution by minimizing $||I(\theta)-\sum T(\theta,\lambda_n)  I(\lambda_n)||^2$. The reconstructed spectra (red dashed curve with circular markers) match well with the input spectra (black solid curve) set in the simulation (Figures \ref{Fig:Figure4_LED_Intensity2Pos_simulation}j-l).

\begin{figure}[t]
\centering\includegraphics[width=7cm]{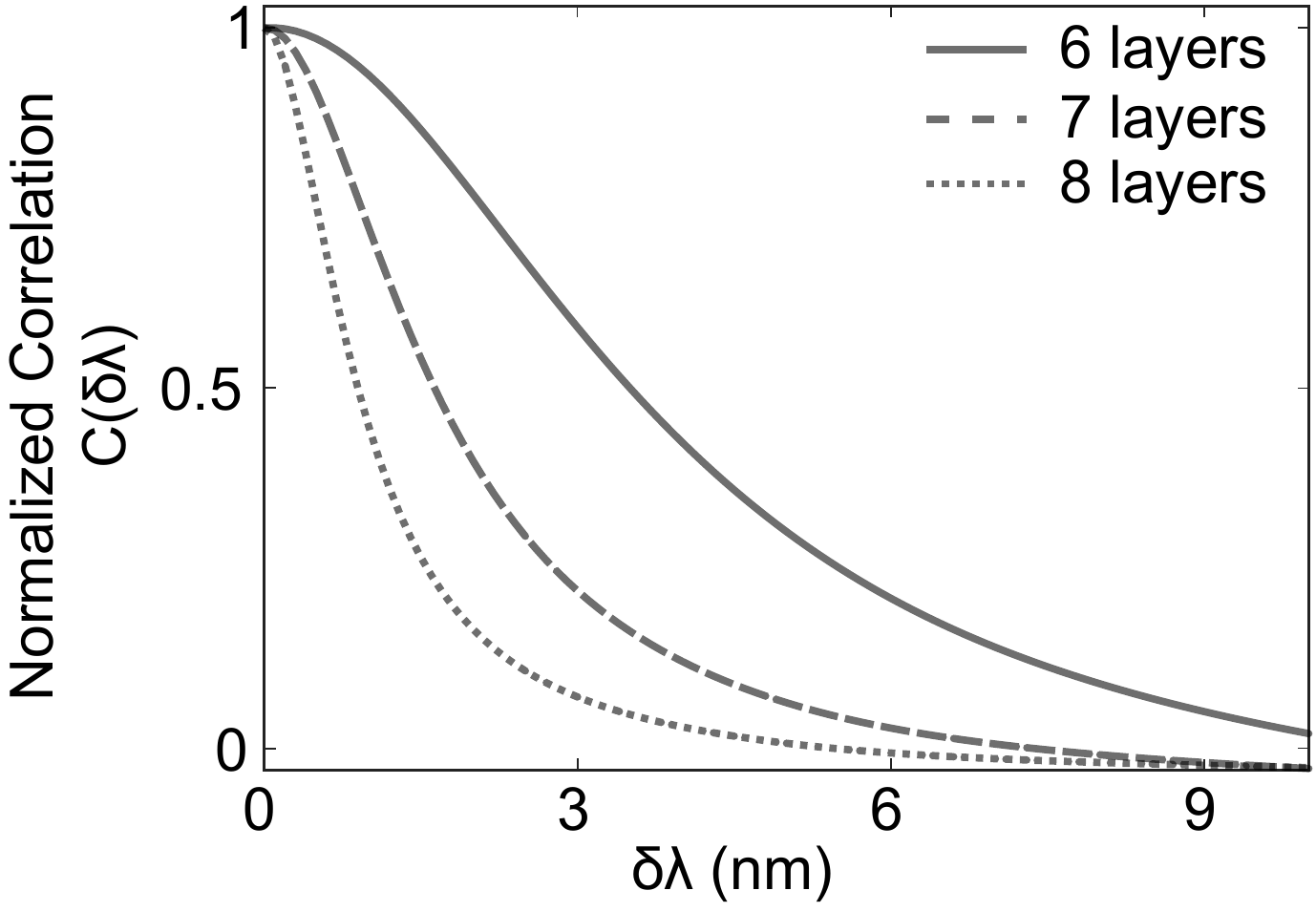}
\caption{\textbf{Spectral Correlation}: Calculated spectral correlation function for five, six, and seven periodic layers in each DBR.} 
\label{Fig:Fig_SpectralCorrelation}
\end{figure}

\textit{Spectral Resolution:} We determine the spectral resolution of our system by computing the spectral correlation function \cite{redding2013, Redding2016}:
\begin{equation}
\label{eqn:SpectralCorrelation}
C(\delta\lambda) = \Bigg \langle \frac {\langle T(\theta,\lambda)  T(\theta,\lambda+\delta\lambda) \rangle_\lambda}{ \langle T(\theta,\lambda)\rangle_\lambda \langle T(\theta,\lambda+\delta\lambda)\rangle_\lambda} - 1 \Bigg \rangle_\theta
\end{equation}
where $\langle$ $\rangle$ represents the average over $\lambda$ or $\theta$ as denoted by the subscript. Figure \ref{Fig:Fig_SpectralCorrelation} shows calculated spectral correlation function in the wavelength space. The half-width at half-maximum represents the wavelength separation between 50$\%$ uncorrelated transmitted intensities and is considered as the spectral resolution of the system \cite{redding2013, redding2013all, wang2014}. The calculated spectral resolution is $\approx$3 nm for our designed photonic crystal cavity that consists of six periodic layers in each DBR (solid gray curve in Figure \ref{Fig:Fig_SpectralCorrelation}). We note that the calculated spectral resolution value and the cavity linewidth are similar. The spectral resolution can be improved by increasing the number of periodic layers in each DBR: $\approx$2 nm and 1 nm for seven and eight periodic layers, respectively (dashed gray and dotted gray curves in Figure \ref{Fig:Fig_SpectralCorrelation}). 

\begin{figure}[t]
\centering\includegraphics[width=7cm]{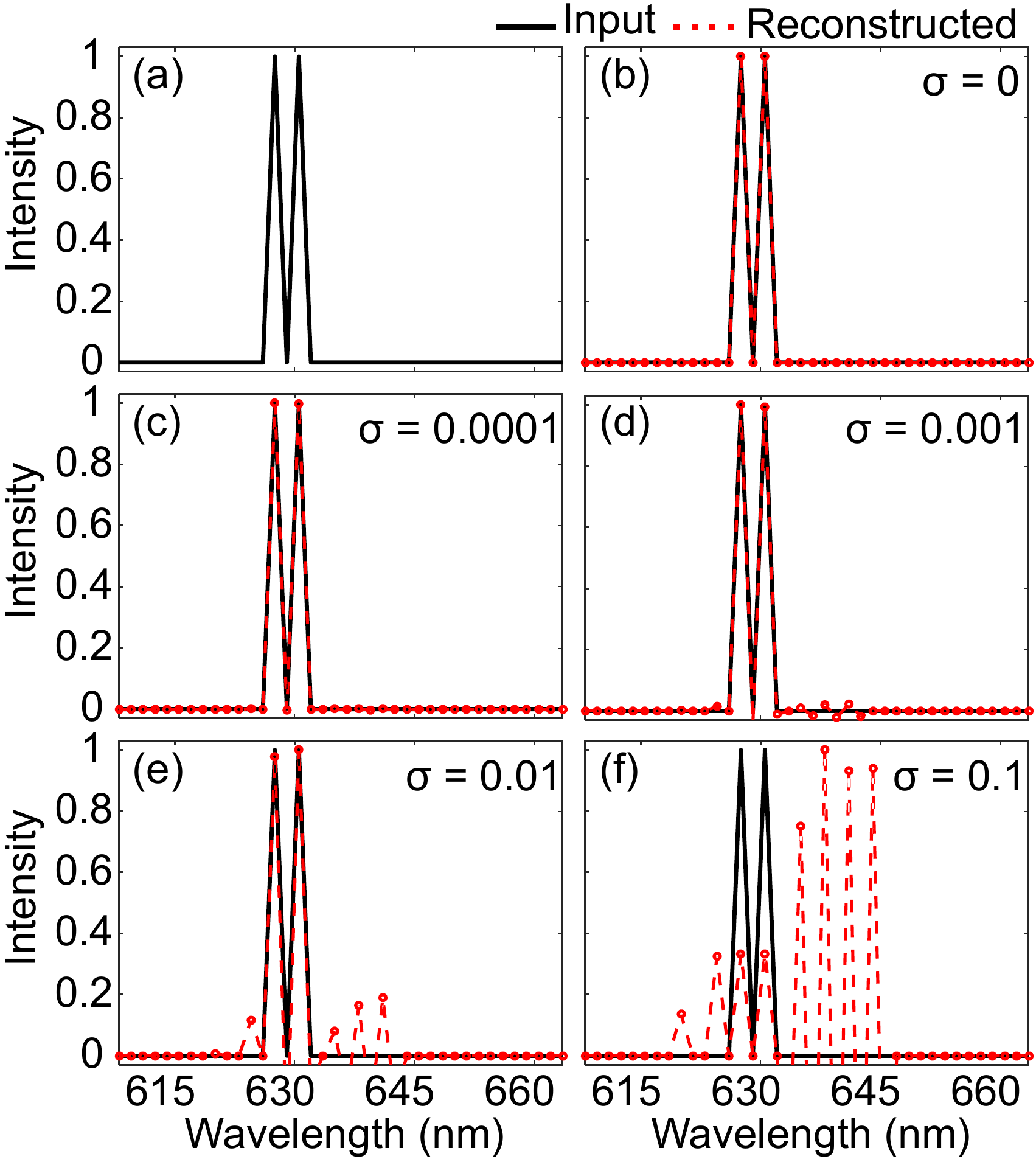}
\caption{\textbf{Spectral reconstruction in the presence of error}: (a) Input spectrum as two resolution-limited peaks separated by 3 nm. (b-f) Reconstructed spectra for $\sigma=$ 0, 0.0001, 0.001, 0.01, and 0.1. The black solid curve represents the input spectrum. The red dashed curve with open circles represents the reconstructed spectrum.} 
\label{Fig:Fig_Spectrum_with_error}
\end{figure}

The spectral resolution also gets affected by measurement errors and instrument noise. To estimate robustness of our reconstruction algorithm, we reconstruct an input spectrum, comprising of two resolution-limited spectral peaks separated by 3 nm, by introducing error in the transmitted spatial intensity profile as following\cite{bao2015}:
\begin{equation}
\label{eqn:TransferFunction_error}
I(\theta)(1+\epsilon_{rand,\theta}) =  \sum_{n=1}^{N} T(\theta,\lambda_n) * I(\lambda_n)
\end{equation}
where $\epsilon_{rand,\theta}$ is a random number for each $\theta$ sampled from a Gaussian distribution with mean zero and standard deviation $\sigma=$ 0, 0.0001, 0.001, 0.01, and 0.1, representing different levels of error. Figure \ref{Fig:Fig_Spectrum_with_error}a shows the input spectrum sampled at a wavelength interval of 1.5 nm. For $\sigma=$ 0 (Figure \ref{Fig:Fig_Spectrum_with_error}b), the reconstructed spectrum exactly matches with the input spectrum (similiar to the figures \ref{Fig:Figure4_LED_Intensity2Pos_simulation}j-l). By increasing $\sigma$ (Figures \ref{Fig:Fig_Spectrum_with_error}c-f), we add more error to the spatial intensity profile, and we observe that our system is able to successfully reconstruct the location of the two peaks up to an error level $\sigma=$ 0.01. The least square error between the reconstructed and the input spectrum, averaged over 100 simulations, is 0.00011, 0.0107, 1.1926, and 6.0831  for $\sigma =$ 0.0001, 0.001, 0.01, and 0.1, respectively.

In the next section, we discuss experiments that provide proof-of-concept demonstration of our designed platform. 

\section*{Proof-of-concept Demonstration}

\begin{figure}[b]
\centering\includegraphics[width=7cm]{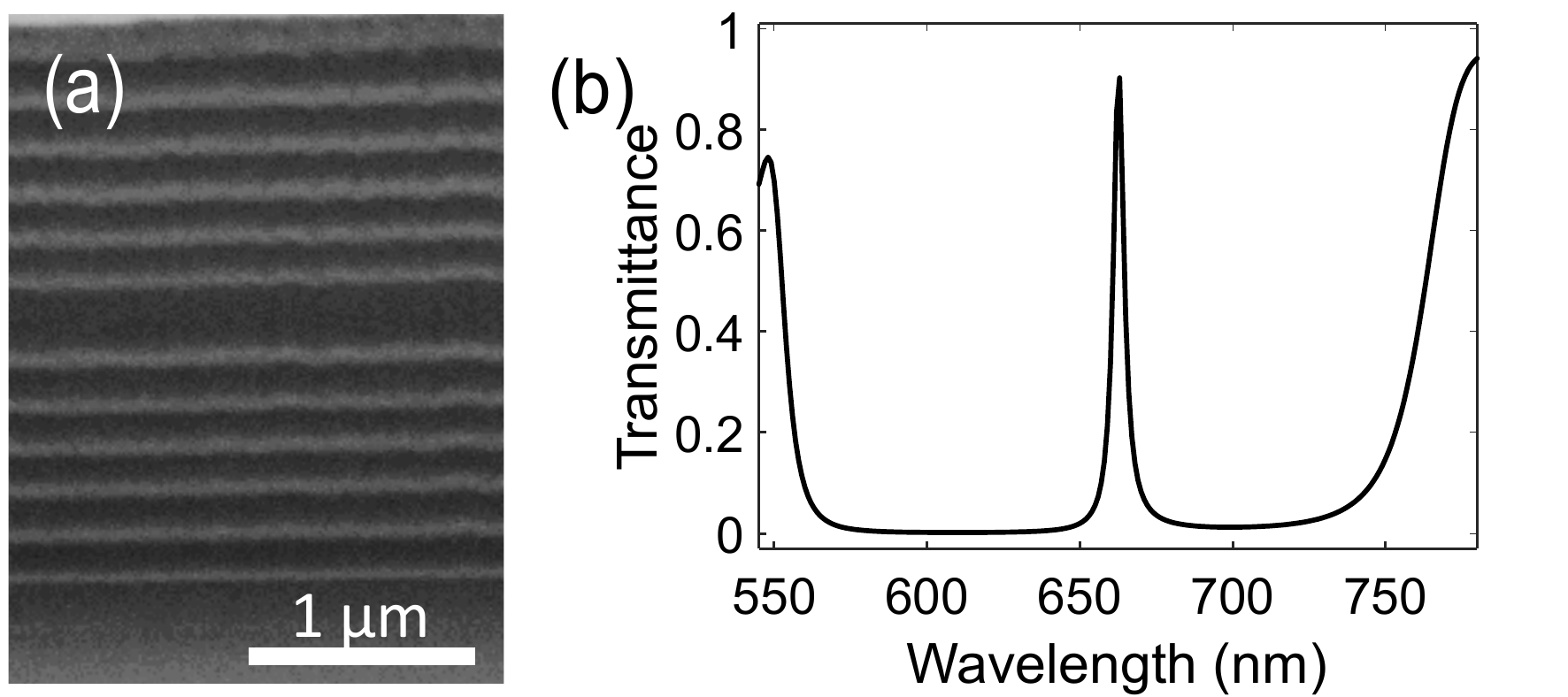}
\caption{\textbf{Characterization of a fabricated photonic crystal cavity}: (a) A representative scanning electron microscope image of a one-dimensional photonic crystal cavity. (b) Transmission spectrum of the fabricated photonic crystal cavity.} 
\label{Fig:Fig5_SEMandTransmission}
\end{figure}

\begin{figure*}[t]
\centering\includegraphics[width=14 cm]{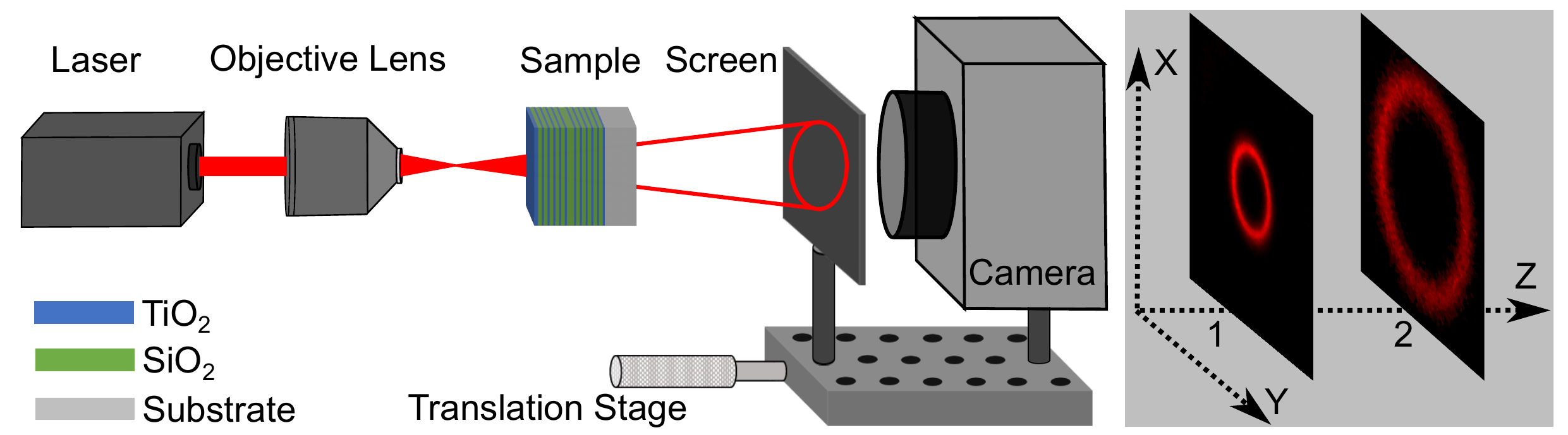}
\caption{\textbf{Optical setup}: Schematic of the optical setup. Translation stage helps to capture the intensity profiles at different positions} 
\label{Fig:Fig6_Optical_Setup}
\end{figure*}

\textit{Sample fabrication:} Figure \ref{Fig:Fig5_SEMandTransmission}a shows the scanning electron microscope image of a photonic crystal cavity fabricated using radio-frequency magnetron sputtering on a 1 inch$^2$ glass substrate. The structure consists of three parts: lower DBR, upper DBR, and a cavity. The  DBR consists of six alternating thin films of SiO$_2$ and TiO$_2$  with average thicknesses of ${109 \pm 1}$ nm  and  ${77 \pm 1}$ nm and refractive indices of 1.46 and 2.05, respectively. Figure \ref{Fig:Fig5_SEMandTransmission}b shows the transmission spectrum the fabricated cavity with resonance at 663 nm  and stopband from 550-765 nm. The transmission spectrum is estimated, using the transfer matrix method, from the experimentally measured reflection spectrum at 8$^\circ$.

\textit{Optical setup:} We use two linearly polarized (along X) and collimated laser beams (Gaussian profile in real space) to characterize our fabricated photonic crystal cavities. The laser wavelengths are 632.9 nm and 656.3 nm, and the typical optical power is  $\approx$ 200 $\micro$W. We focus the laser beam using a 0.9 NA objective lens and capture the transmitted beam profiles at two positions (separated by 33 mm) along the propagation direction (along Z) using a screen and a web camera, both mounted on a translation stage (Figure \ref{Fig:Fig6_Optical_Setup}).

\begin{figure}[t]
\centering\includegraphics[width=7 cm]{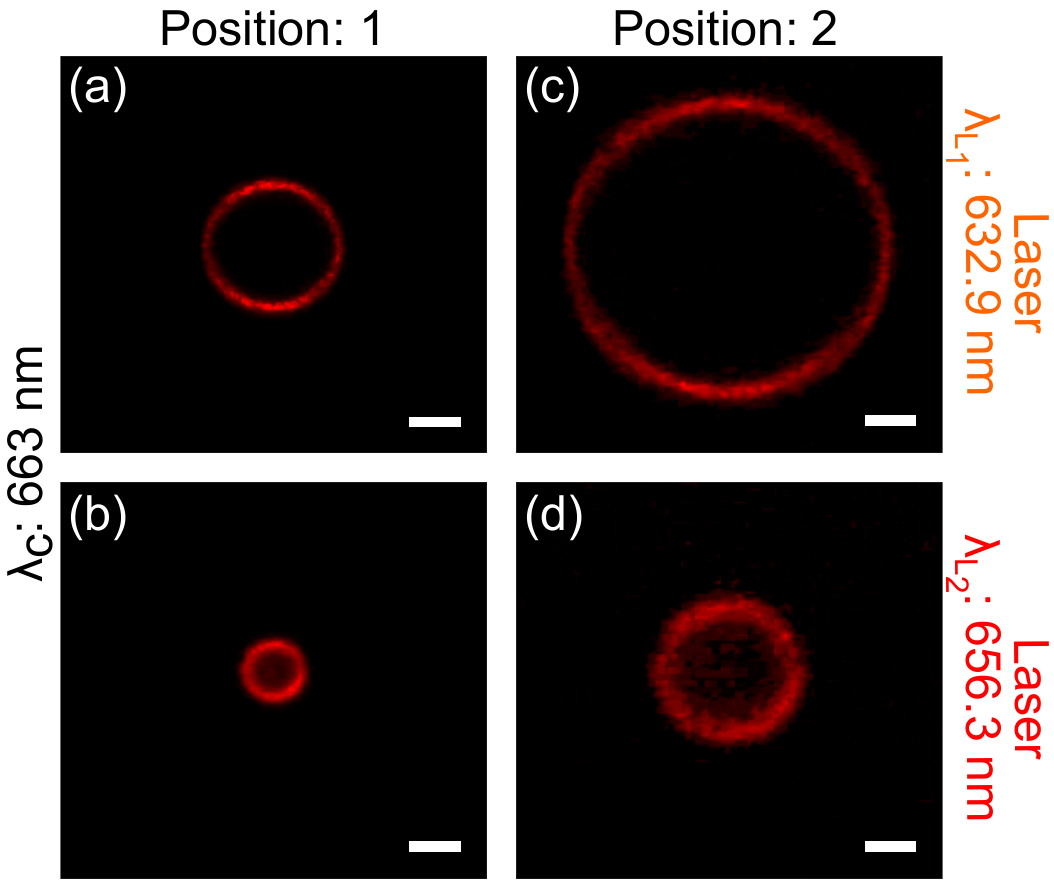}
\caption{\textbf{Experimentally measured transmitted intensity profiles of lasers}: Transmitted intensity profiles of laser beam after passing through the sample at (a)-(b) position 1 and (c)-(d) position 2. Laser wavelengths are 632.9nm, and 656.3nm. The distance between on-axis position 1 and position 2 is 33 mm. All scale bars are 10 mm. All images of the transmitted beam are adjusted to the same maximum brightness and the same contrast enhancement.}
\label{Fig:Fig7_Laser_Annular_profile}
\end{figure}

\begin{figure}[t]
\centering\includegraphics[width=7cm]{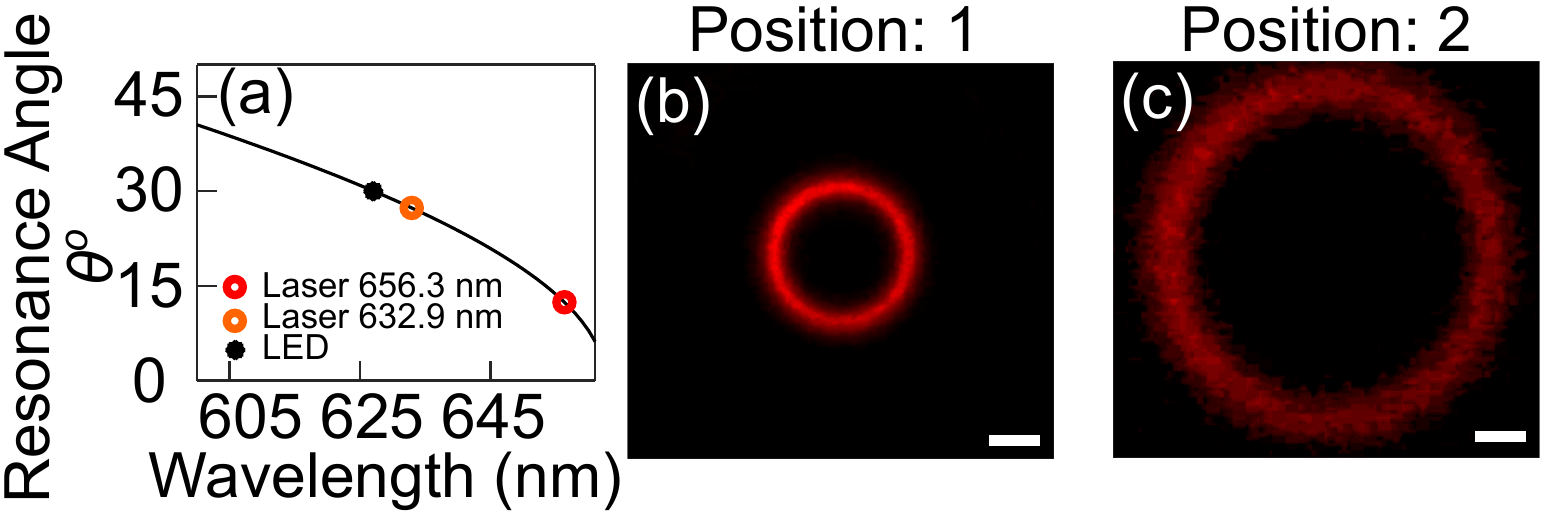}
\caption{\textbf{Experimentally calculated dispersion relation and measured transmitted intensity profiles of LED}:(a)Averaged resonance angle as a function of  input wavelength. (b)-(c) Measured Transmitted intensity profiles of a LED after passing through the photonic crystal cavity at position 1 and position 2. The distance between on-axis position 1 and position 2 is 33 mm. All scale bars are 10 mm. All images of the transmitted beam are adjusted to the same maximum brightness and the same contrast enhancement.}%
\label{Fig:Fig8_Exp_LED_Profiles}
\end{figure}

\textit{Generating dispersion relation for fabricated cavity:} We characterize the fabricated cavity by capturing the transmitted intensity profiles at two positions along the propagation direction for the two laser wavelengths (Figure \ref{Fig:Fig7_Laser_Annular_profile}). We observe similar behavior in experiments as in the simulations (Figure \ref{Fig:Fig2_System_Design}).  We calculate the dispersion relation, for both TE and TM polarization directions, by estimating resonance angles for the two laser wavelengths from the diameters of the annular transmitted beams (Figures \ref{Fig:Fig7_Laser_Annular_profile}a-d) and using eq. \ref{eqn:Lambda_Theta}. Averaging over the two polarizations, we plot the dispersion relation in  figure \ref{Fig:Fig8_Exp_LED_Profiles}a as the solid curve and represent the average resonance angles corresponding to the two laser wavelengths with open circles. Using the dispersion relation, we estimate the cavity resonance wavelength to be 662.68 nm and the cavity linewidth to be 4.61 nm (both at normal incidence), which match well with the measured values of 663 nm and 4.24 nm using a commercial spectrophotometer. We attribute the difference in simulated and experimentally observed values of the cavity linewidth to fabrication imperfections. Unavailability of a tuneable laser source limits us from experimentally estimating the transfer function of the system. However, we can use the experimentally estimated dispersion relation to determine center wavelength and linewidth of a narrowband spectrum. In the next section, we will use the fabricated sample to characterize a LED source. 

\textit{Characterizing an unknown light spectrum}: We replace the laser with a LED in the optical setup (Figure \ref{Fig:Fig6_Optical_Setup}) and capture the transmitted intensity profiles at two different positions (Figures \ref{Fig:Fig8_Exp_LED_Profiles}b-c). We observe that the intensity profiles are symmetric in both the vertical and horizontal directions, indicating the LED source is unpolarized. We calculate the resonance angle to be 29.97$^\circ$ (represented by a star marker in figure \ref{Fig:Fig8_Exp_LED_Profiles}a) and the resonance width to be 7.34$^\circ$ from the transmitted intensity profiles. We estimate the center wavelength to be 627.02 nm and the linewidth to be 16.72 nm for the LED source using the dispersion curve (Figure \ref{Fig:Fig8_Exp_LED_Profiles}a). These estimated values are close to the values measured using a commercial spectrometer with a resolution of 1 nm (Supplementary Figure \ref{SuppFig: LEDspec}).

\begin{figure*}[t]
\centering
\includegraphics[width=14 cm]{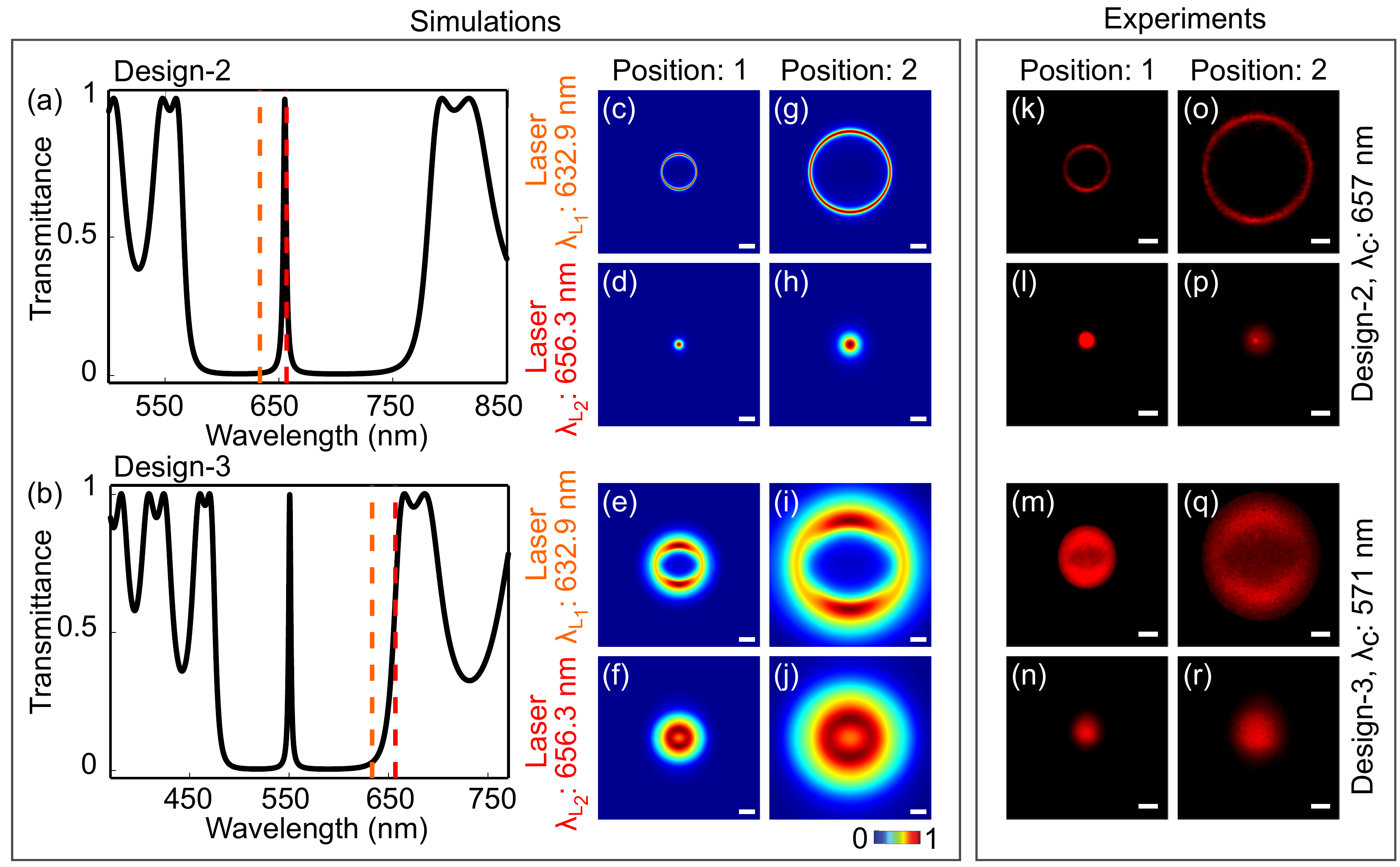}
\caption{(a-b) Simulated transmittance spectra of two more photonic crystal cavities (Design-2 and Design-3) with different resonance wavelengths and stopband ranges. The vertical dashed lines represent the laser wavelengths, 656.3 nm and 632.9 nm, used for characterization. Simulated far-field intensity profiles of laser beams after passing through the samples at (c)-(f) position 1 and (g)-(j) position 2. Experimentally measured transmitted intensity profiles of laser beam after passing through the fabricated samples at (k)-(n) position 1 and (o)-(r) position 2. The distance between on-axis position 1 and position 2 is 33 mm. All scale bars are 10 mm. Intensity of each panel is individually normalized.  Design-2 and Design-3 show an annular profile for only laser wavelength $\lambda_{L1}$. The width of the annular profile of Design-3 is broader than Design-2. While the width of the annular beam in Design-2 (and also the design discussed in the main manuscript) is determined by the linewidth of the cavity resonance, the width of the annular beam in Design-3 is determined by the bandedge of the stopband and the NA of the lens. In Design-3, $\lambda_{L1}$ is slightly shorter than bandedge of the stopband, and therefore, transmits at an oblique angle of incidence. In this case, the inner diameter of the annular beam is decided by the smallest angle of incidence required for non-zero transmission, which is a function of the difference between the wavelengths of the incident laser and the band-edge, and the outer diameter of the annular beam is decided by the largest angle of incidence, which is a function of the NA of the lens. The laser wavelength $\lambda_{L2}$ is outside the stopband, and therefore, annular profile is not observed for Design-3.}
\label{SuppFig: design2and3}
\end{figure*}

\begin{figure}[b]
\centering\includegraphics[width=7 cm]{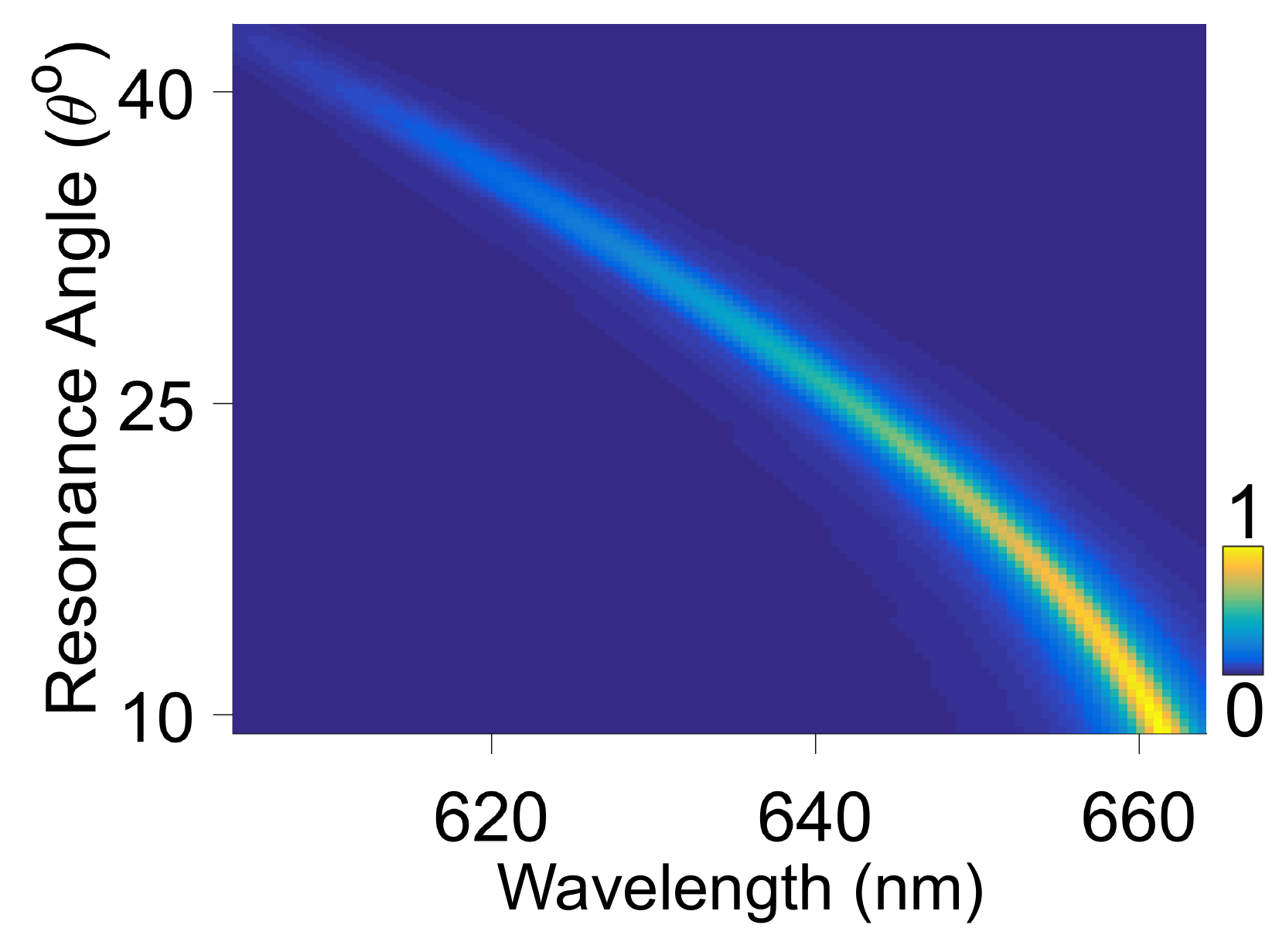}
\caption{Simulated transfer function of the system, averaged over TM and TE polarization directions.}
\label{SuppFig: Transferfunction}
\end{figure}

\begin{figure}[b]
\centering\includegraphics[width=7 cm]{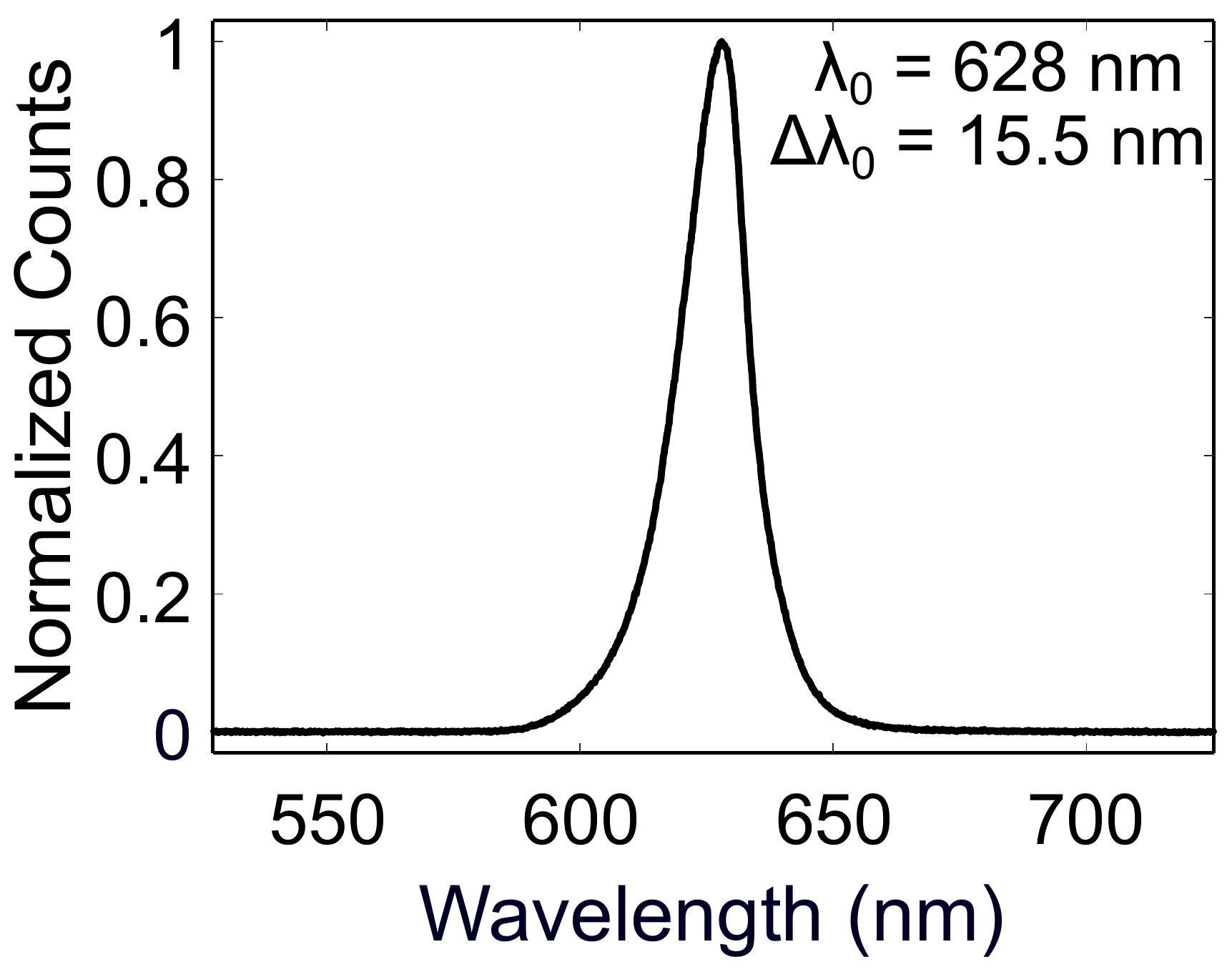}
\caption{LED spectrum measured using a commercial spectrometer at a resolution of 1 nm.}
\label{SuppFig: LEDspec}
\end{figure}

In conclusion, we have demonstrated a reconstructive spectrometer using a one-dimensional photonic crystal cavity as a dispersive element. The planar architecture of the cavity makes the spatial profile of the  transmitted intensity, and thus, the reconstruction of the spectrum invariant to translation of the photonic crystal cavity in the transverse or the longitudinal plane. Our portable, economical, and robust-to-misalignment spectrometry platform fills the gap between the large benchtop spectrometers that suffer from size-resolution tradeoff and the on-chip spectrometers that require complex fabrication techniques. The spectral resolution of our system is governed by the cavity linewidth and the spectral range is governed by the stopband, both of which can be tuned by tailoring the photonic crystal parameters (Supplementary Figure \ref{SuppFig: design2and3}). The light sensitivity of the system can be improved by replacing the screen and the web camera with an electronic sensor. Apart from spectrometry, our technique can also be employed for characterizing photonic crystal cavities with just two known laser sources using the dispersion relation.

\begin{acknowledgments}
We acknowledge funding support from SERB (SB/S2/RJN-134/2014, EMR/2016/007113), DST (SR/FST/ETII-072/2016), and IMPRINT-I (no. 4194) project. We thank R. Vijaya for fabrication support. We also acknowledge support of the FIB facility of MEMS department at IIT Bombay. 
\end{acknowledgments}

\bibliography{References}

\beginsupplement
\pagebreak
\newpage

\end{document}